# Thermodynamic Fluctuations in Magnetic Phase Transition: Invar as a Prototype

Y. Wang, S. L. Shang, H. Zhang, L. -Q. Chen, and Z. -K. Liu

Materials Science and Engineering, The Pennsylvania State University, University Park, PA 16802, USA.

We propose a first-principles formulism for system with spin fluctuations and apply it to the ordered $Fe_3Pt$ to uncover the Invar anomalies, including negative thermal expansion and spontaneous magnetization. The theory has coherently predicted the finite temperature intermixing between the fully ferromagnetic configuration and the spin-flipping configurations. We also discover a tri-critical point at which a high-temperature second-order phase transition, between the fully ferromagnetic configuration and the spin-flipping configurations, becomes first-order at low temperatures.

The concise description of the thermodynamic fluctuations among many electronic states, each distinguished by the spin configurations, is one of the most enigmatic aspects of modern condensed matter physics. Solution of this can reveal the microscopic origin of the intriguing properties of many materials. The fundamental examples are the elemental metals Fe, Co, and Ni, and Invar alloys [1, 2], which undergoes the well-documented ferromagnetic-paramagnetic transition at its Curie temperature [3]. Further examples include the cuprates [4], the newly-discovered $LaO_{1-x}F_xFeAs$ [5], and the heavy-fermion metals [6], most of which follow the antiferromagnetism → spin glass → supercoductor phase transition sequence. More complicated examples are $ZrZn_2$ [7] and $\varepsilon$–Fe [8] for which superconductivity and ferromagnetism have been found to coexist as well as multiferroics [9] and lithium transition metal phosphates for rechargeable batteries [10] for which ferroelectric, ferro/antiferromagnetic, and ferroelastic phases coexist.

In this Letter, we develop a first-principles theory to a system with thermodynamic fluctuations among many spin configurations. The present application is to $Fe_3Pt$ – one representative of a family of materials called Invar which exhibits a broad variety of characteristic anomalies [2], best known for exceptionally low or even negative coefficients of thermal expansion over a wide temperature range.

Invar was discovered in intermetallic $Fe_{65}Ni_{35}$ alloy in 1897 by Guillaume [11] who received a Nobel Prize in Physics for the discovery in 1920. Despite extensive theoretical and experimental activities [2, 12-29] over the last century, stimulated by their wide-spread applications in scientific instruments, there is a lack of a microscopic understanding that can satisfactorily explain all the Invar anomalies. The pre-existing theories include: i) the Weiss 2-$\gamma$ model [15, 19, 24], ii) the non-collinear spin model [20, 25], and iii) the disordered local moment (DLM) approach [27, 30-32].

At 0 K, we assume that in addition to the fully ferromagnetic configuration (FMC) where all the spins on each Fe atom line up along one direction, there exist many other spin-flipping configurations (SFC) with a fraction of the spins in the opposite direction. Above 0 K, we postulate that each spin configuration has its own characteristic thermal vibration and thermal electronic excitation. We then propose that FMC and various SFCs coexist and their thermal populations, dictated by their individual Helmholtz energy, are both temperature and volume dependent. We emphasize that our formulism is fundamentally different from the existed models. We first calculate the free energies of each spin configuration as a function of temperature and volume independently, and then mix representative spin configurations through statistical analysis at finite temperatures.

Let us consider a lattice with $N$ atoms under the constant volume $V$ and temperature $T$. We start from the partition function of a specific spin configuration $\sigma$, which is known as [33]:

$$Z^{\sigma} = \sum_{i \in \sigma, \rho \in \sigma} \exp[-\beta \varepsilon_i(N,V,\rho)] = \exp[-\beta F^{\sigma}(N,V,T)], \tag{1}$$

where $\beta = 1/k_B T$, $i$ identifies all the vibrational states within $\sigma$, $\rho$ labels the electronic distributions within $\sigma$, $\varepsilon_i(N,V,\rho)$ is the eigenvalue of the corresponding microscopic Hamiltonian associated with $\sigma$, and $F^{\sigma}(N,V,T)$ is its Helmholtz energy.

Then the partition function for a system with multi spin configurations, $Z$, can be written as:

$$Z = \sum_{\sigma} w^{\sigma} Z^{\sigma}, \tag{2}$$

where $w^\sigma$ is the multiplicity of the spin configuration $\sigma$. It is immediately apparent that $x^\sigma = w^\sigma Z^\sigma / Z$ is the thermal population of the spin configuration $\sigma$. Furthermore, with $F = -k_B T \ln Z$ [3], we obtain the Helmholtz energy of the system as

$$F(N,V,T) = -k_B T \sum_\sigma x^\sigma \ln Z^\sigma + k_B T \left[ \sum_\sigma x^\sigma \ln Z^\sigma - x^\sigma \ln Z \right]$$
$$= \sum_\sigma x^\sigma F^\sigma(N,V,T) + k_B T \sum_\sigma x^\sigma \ln(x^\sigma / w^\sigma). \quad (3)$$

Equation (3) relates the total Helmholtz energy of a system with many spin configurations, $F(N,V,T)$, and the Helmholtz energies of individual spin configurations, $F^\sigma(N,V,T)$. An important result of Eq. (3) is the configurational entropy of multi spin configurations

$$S_f(N,V,T) = -k_B \sum_\sigma w^\sigma \left[ x^\sigma / w^\sigma \ln(x^\sigma / w^\sigma) \right]. \quad (4)$$

We use the SFCs derived [34] from a system with 12-atom 3×1×1 supercell. We consider contributions to $F^\sigma$ from three resources: i) 0 K electronic energy; ii) the lattice vibration; and iii) thermal electronic excitation. To calculate the 0 K energy, we have employed the VASP package within the projector-augmented wave (PAW) method [35, 36]. The exchange-correlation part of the density functional was treated within the GGA of Perdew-Burke-Ernzerhof (PBE) with the interpolation formula of Vosko et al. [37]. For all SFCs, the local structures are relaxed. For the evaluation of the thermal electronic excitation, we have employed integration over the electronic density-of-states through Fermi-Dirac distribution [38]. For the lattice vibration, we find that the Debye-Grüneisen approach [39, 40] is a fast and yet accurate enough solution.

Fig. 1 presents the first-principles 0 K total energies of 36 non-equivalent SFCs as well as the FMC as a function of atomic volume. We indeed find a number of SFCs, whose energies are in the range of ~1 mRy/atom to that of the FMC. It is interesting to note that all the SFCs studied herein have the equilibrium averaged atomic volumes at least 1.8 % smaller than that of the FMC, the 0 GPa ground state. The present calculation shows that the nonmagnetic configuration (not shown in Fig. 1) has a very small atomic volume of 11.66 Å$^3$/atom, and its energy is higher than both FMC and all SFCs.

Based on the free energy dependency on temperature and volume, we have calculated the T-V phase diagram that is plotted in Fig. 2(a). It clearly shows a tri-critical point at 141 K and $V$ = 12.61 Å$^3$ with $P$ = 5.81 GPa. Below the tri-critical point, it is a two-phase miscibility gap (the shadow area enclosed by the dotted lines). Above the tri-critical point, the phase transitions between FMC (at large volumes) and SFCs (at small volumes) are of second-order in nature (the transition volumes are determined by the condition that $F_{SFCs}$, the free energy of all SFCs, equals to $F_{FMC}$, the free energy of FMC). The existence of such a tri-critical point is supported by experimental measurements [18, 28, 41]. For example, Abd-Elmeguid and Micklitz [18] observed a critical point at ~110 K and 6.0 GPa, similar to the values of ~130 K and 7.0 GPa obtained by Matsushita et al. [28]. We also provide the *T-P* phase diagram (Fig. 2(b)) showing the phase boundary between FMC and SFCs, where the data points are the measured pressure dependence of the Curie temperature ($T_c$) [18, 28]. The agreement between the measurements and our predictions is remarkable. We want to add that the configuration mixing considered in our mode through Eq. (3) plays key role in predicting the existence of the tri-critical point. For demonstration, we have also calculated phase boundary between FMC and SFCs without considering the configuration mixing between the two phases. It is seen that the phase transition between FMC and SFCs is always of first-order. It should be pointed out that, the classical Weiss 2-γ model [15] predicts only first-order phase transitions while the non-collinear spin model yields only second-order phase transitions at all temperatures [25] as pointed out by Nataf et al. [41].

We illustrate the predicted thermal volume expansion in Fig. 3a and the derived linear thermal expansion coefficient in Fig 3b. For comparison, we also include the available experimental data for $Fe_3Pt$ [17] and $Fe_{72}Pt_{28}$ [16, 21]. We predicted a positive thermal expansion from 100 K to 288 K, followed by a negative thermal expansion in the range of 289 ~ 449 K, and then a positive thermal expansion again at >450 K, in excellent agreement with experiments [16, 17, 21]. The only disagreement between our calculations and experiments occur at T < 100 K where the calculations did not reproduce the negative thermal expansion for $Fe_3Pt$. Large supercell may be necessary for low temperature.

To fully understand Invar, an attempt is made to develop a formulation to calculate its spontaneous magnetization, $M_s(T)$. The formulation of $M_s(T)$ of Invar has been enduring challenges as both the Bloch $T^{3/2}$ and the Stoner $T^2$ laws [3] failed to describe the magnetic moment dependence on temperature of Invar. Maruyama and coworkers [2, 42-44] proposed a fitting formula by combing the spin-wave excitations and a second excitation whose physical nature was unknown.

We postulate that $M_s(T)$ of Invar is a thermal average over the spontaneous magnetization of the individual spin configuration as:

$$M_s(T) = \sum_\sigma x^\sigma M_s^\sigma(T), \tag{5}$$

with that the spontaneous magnetization of the FMC, $M_{SW}^{FMC}(T)$, obeys the spin-wave theory with the Bloch $T^{3/2}$ form [3] and the spontaneous magnetization of the SFC, $M_{MF}^{SFC}(T)$, obeys the mean-field theory with the Brillouin expression [3].

The calculated spontaneous magnetization and thermal populations of FMC and that of the sum over all SFCs vs. temperature curves are plotted in Fig. 4a and 4b. Our calculated $M_s(T)$ demonstrates several important physics for Invar:

1) At low temperature ($T/T_c < 0.5$), $M_s(T)$ is completely dictated by the FMC (Fig. 4a), as it is seen that $x^{FMC}$ takes its maximum value of 1.0 for $T/T_c < 0.5$ (Fig. 4b).

2) For $T/T_c > 0.5$, Fig. 4b shows that $x^{FMC}$ decreases in an exponential form which results in a 'tail' on $M_s(T)$ around $T_c$ = 460 K (Fig. 4a), in agreement with the experiments [2, 44].

In summary, through explicitly considering the freedom of spin in partition function, we have developed a first-principles formulation of the Helmholtz energy for materials that exhibit thermodynamic fluctuations among different spin configurations. Illustrated with $Fe_3Pt$, the present theory satisfactorily addresses almost of all the observed Invar anomalies.

**Acknowledgements** Calculations were conducted at the LION clusters at the Pennsylvania State University, supported in part by NSF Grants Nos. DMR-9983532, DMR-0122638, DMR-0205232, and DMR-0510180. This research also used resources of the NERSC supported by the Office of Science of the U.S. Department of Energy under the Contract No. DE-AC02-05CH11231. This work was supported in part by a grant of HPC resources from the Arctic

**Figure Captions**

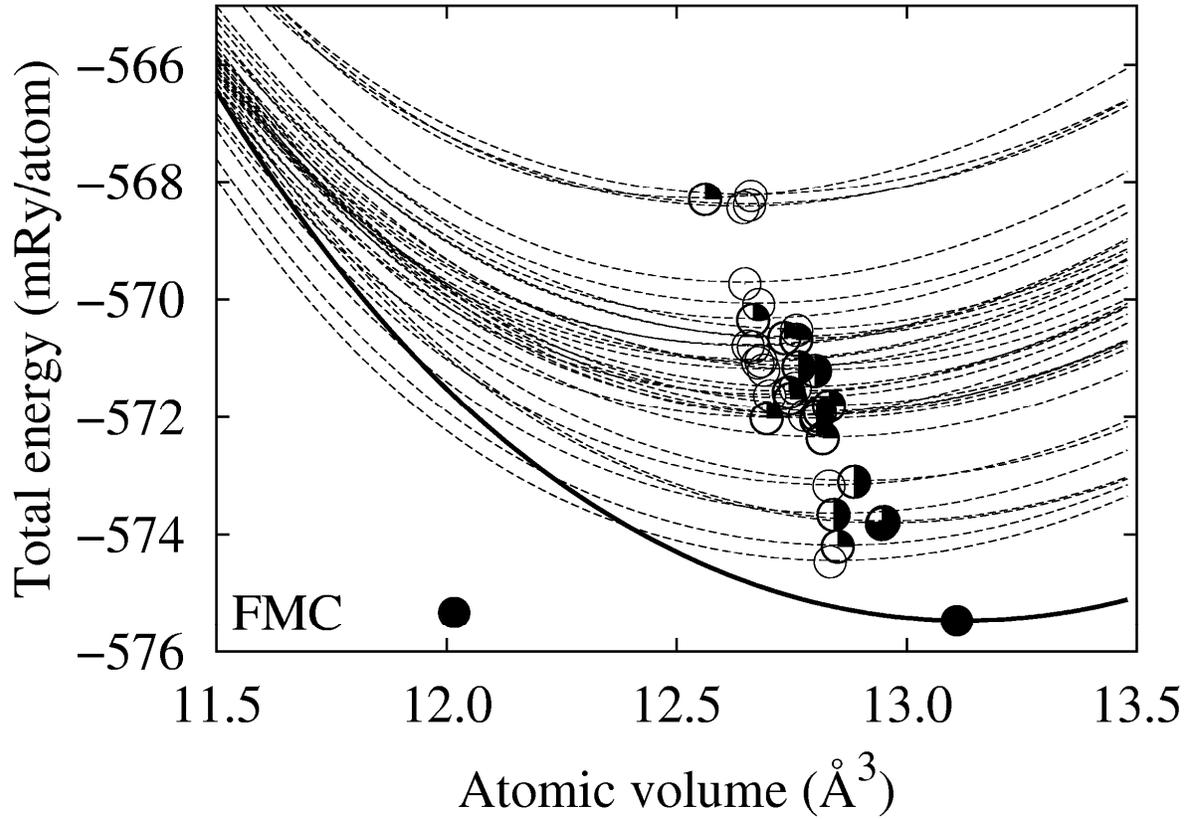

Figure 1. 0 K total energies. The heavy black line represents the FMC. The symbols ○, ◐, ◑, and ◕ with dashed lines indicate the minima of the energy-volume curves of the SFCs with spin polarization rates of 1/9, 3/9, 5/9, and 7/9, respectively.

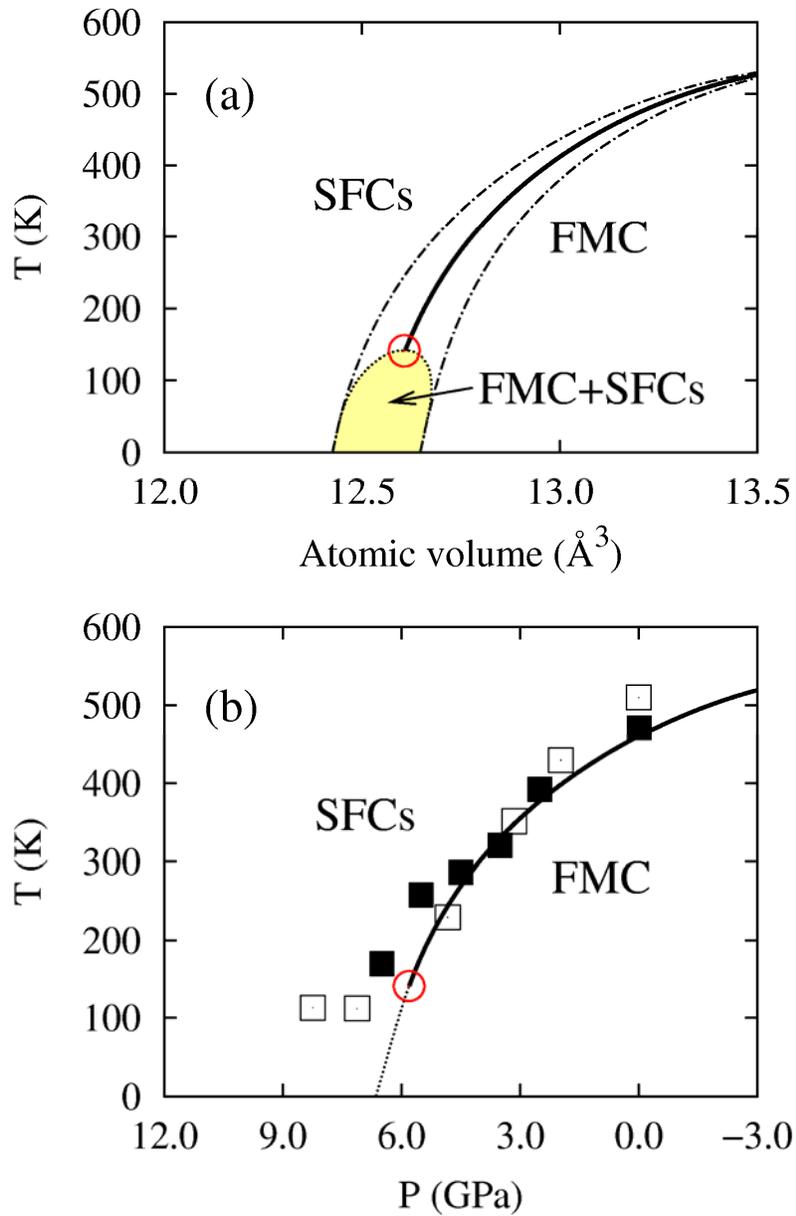

Figure 2 (color online). (a) *T-V* phase diagram of the ordered $Fe_3Pt$. Yellow shadow: the predicted region of two phase mixture; Solid line: the calculated tentative phase boundary above the tri-critical point (○). The dot-dashed lines denote the calculated phase boundary without considering the configuration mixing between FMC and SFCs. (b) *T-P* phase diagram. Dotted line: the calculated phase boundary (assuming no configuration mixing) below the tri-critical temperature; □: Curie temperature ($T_c$) of $Fe_{72}Pt_{28}$, measured by Abd-Elmeguid & Micklitz [18], and ■: $T_c$ of $Fe_{72.8}Pt_{27.2}$, 3measured by Matsushita et al. [28].

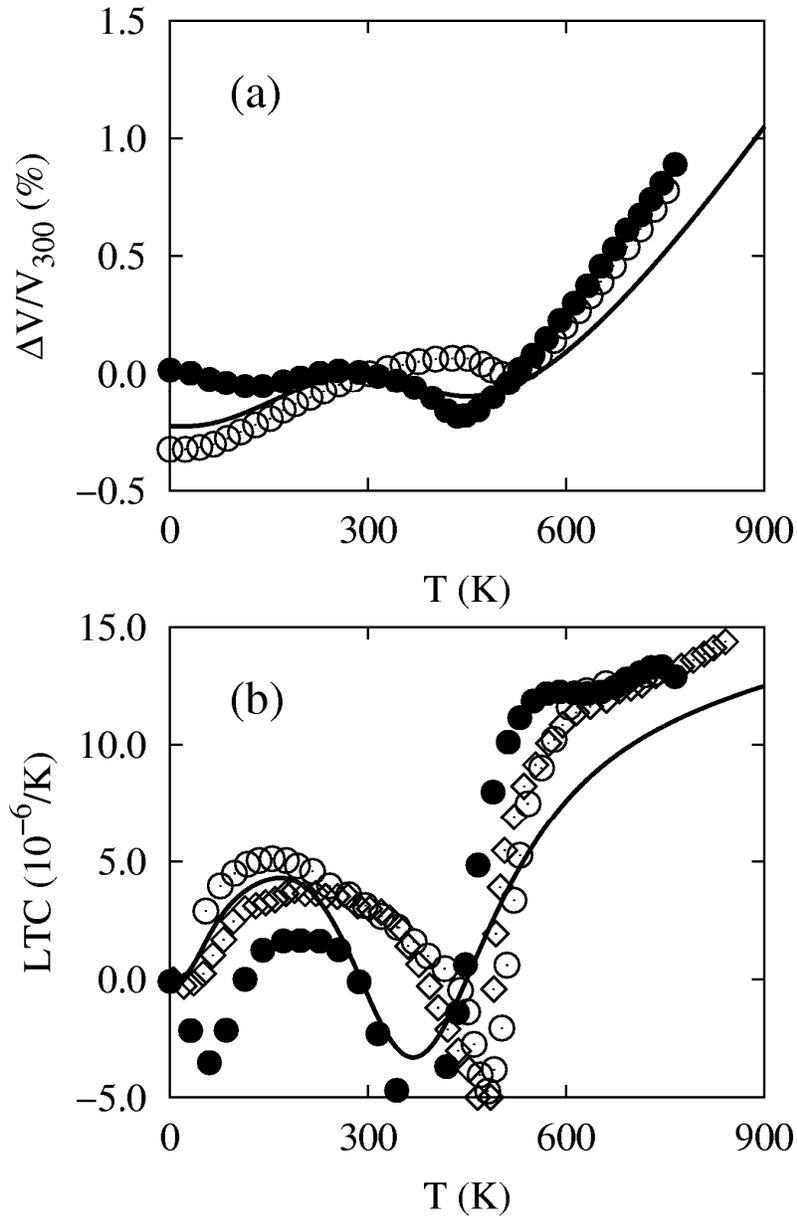

Figure 3. (a) Relative volume increase $(V - V_{300})/V_{300}$ with $V_{300}$ being the equilibrium volume at 300K and 0 GPa for the ordered $Fe_3Pt$. (b) Linear thermal expansion coefficient (LTC). Solid line: the present calculations; ○: $Fe_{72}Pt_{28}$, measured by Sumiyama et al. [16]; ●: $Fe_3Pt$, measured by Sumiyama et al. [17] and ◇: $Fe_{72}Pt_{28}$, shown by Rellinghaus et al. [21]

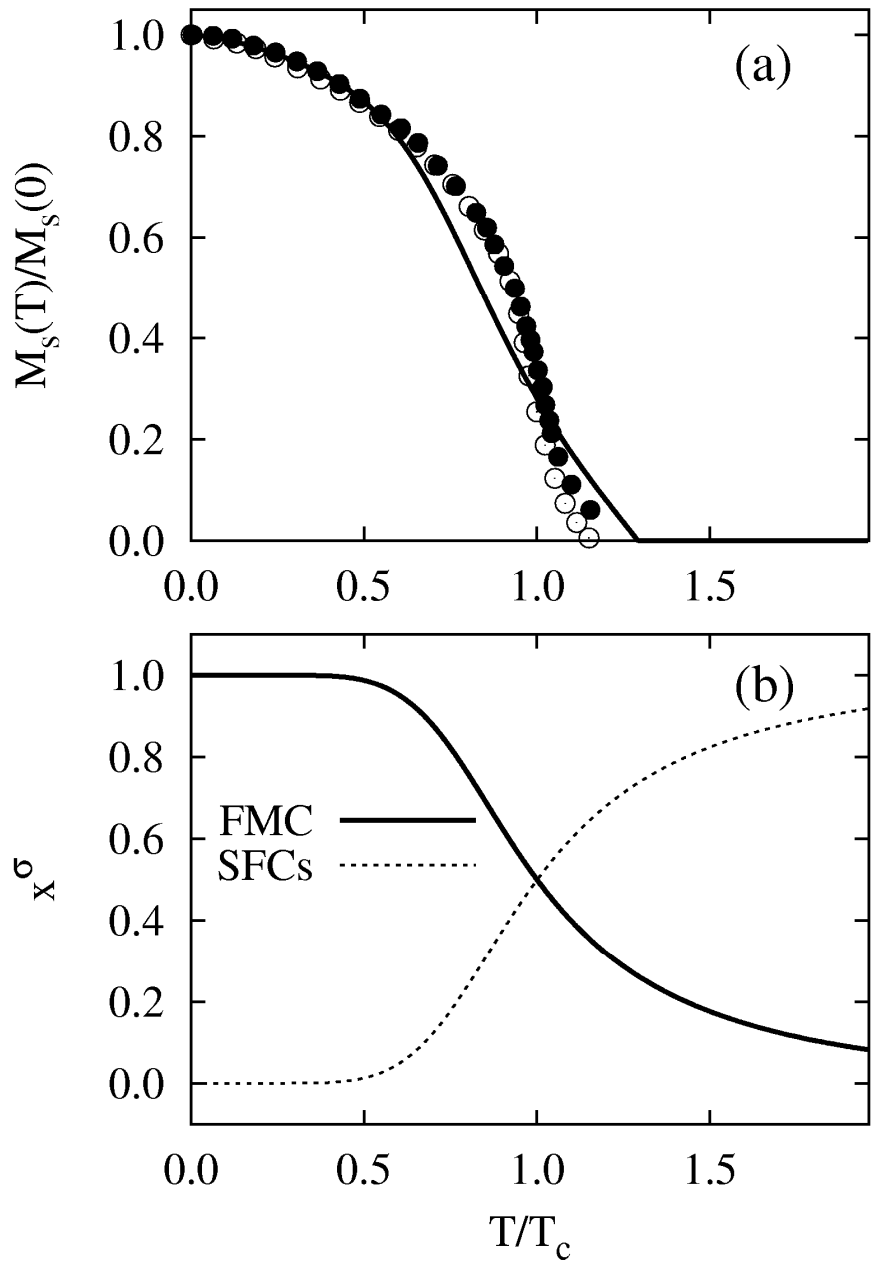

Figure 4. (1) Reduced spontaneous magnetization, $M_s(T)/M_s(0)$, vs. reduced temperature, $T/T_c$. Solid line: the present calculations; ○: $Fe_{72}Pt_{28}$, from the review work by Wasserman [2]; ●: $Fe_{70}Pt_{30}$, measured by Shen et al. [44]. (b) The calculated thermal populations of the FMC (solid line) and that of the sum over all SFCs (dashed line).